\documentclass[aps,prl,preprint,groupaddress, superscriptaddress]{revtex4}
\usepackage[pdftex]{graphicx}
\usepackage{bm}
\usepackage{pifont}
\usepackage{amssymb, amsmath}
\usepackage{dcolumn}
\usepackage{color}

\begin{document}
\title{%
Transport evidence of current-induced nematic Dirac valleys in a parity-time-symmetric antiferromagnet
}
\author{H. Sakai}
\email[Corresponding author: ]{sakai@phys.sci.osaka-u.ac.jp}
\affiliation{Department of Physics, Osaka University, Toyonaka, Osaka 560-0043, Japan}
\author{Y. Miyamoto}
\affiliation{Department of Physics, Osaka University, Toyonaka, Osaka 560-0043, Japan}
\author{M. Kimata}
\affiliation{Institute of Materials Research, Tohoku University, Sendai, Miyagi 980-8577, Japan}
\affiliation{Advanced Science Research Center, Japan Atomic Energy Agency, Tokai, Ibaraki 319-1195, Japan}
\author{H. Watanabe}
\affiliation{Department of Physics, University of Tokyo, Tokyo 113-0033, Japan}
%
\author{Y. Yanase}
\affiliation{Department of Physics, Graduate School of Science, Kyoto University, Kyoto 606-8502, Japan}
\author{M. Ochi}
\affiliation{Department of Physics, Osaka University, Toyonaka, Osaka 560-0043, Japan}
\affiliation{Forefront Research Center, Osaka University, Toyonaka, Osaka 560-0043, Japan}
\author{M. Kondo}
\altaffiliation[Present address: ]{The Institute of Solid State Physics, The University of Tokyo, Kashiwa, Chiba 277-8581, Japan}
\affiliation{Department of Physics, Osaka University, Toyonaka, Osaka 560-0043, Japan}
\author{H. Murakawa}
\affiliation{Department of Physics, Osaka University, Toyonaka, Osaka 560-0043, Japan}
\author{N. Hanasaki}
\affiliation{Department of Physics, Osaka University, Toyonaka, Osaka 560-0043, Japan}
\affiliation{Spintronics Research Network Division, Institute for Open and Transdisciplinary Research Initiatives, Osaka University, Suita, Osaka 565-0871, Japan}
\begin{abstract}
Itinerant antiferromagnets with broken time-reversal symmetry have recently attracted attention, since their spin-split bands enable large magnetotransport responses comparable to ferromagnets despite the negligible spontaneous magnetisation. 
When the inversion symmetry is further broken by the antiferromagnetic order, the emerging odd-parity multipole order renders the bands spin-degenerate but asymmetric in the momentum space.
For such parity-time-symmetric antiferromagnets, it has been predicted that electronic nematicity is induced by current, allowing unconventional nonlinear transport phenomena.
However, their experimental evidence has been lacking.
Here, we report nonreciprocal angular magnetoresistance in the layered Dirac material SrMnBi$_2$ with parity-time-symmetric antiferromagnetic order in its Mn-Bi layers.
By quantitatively modelling the angular and field dependencies using a phenomenological framework, we reveal that the observed nonreciprocal interlayer resistivity arises from the broken four-fold symmetry of the Dirac valleys in the Bi square net adjacent to the Mn-Bi layer.
Furthermore, we demonstrate the alignment of parity-time-symmetric antiferromagnetic domains via current-field cooling, achieving electric-magnetic control of the $f$-wave polarity in momentum space.
The observed switchable nonreciprocal transport associated with current-induced valley symmetry breaking paves the way for novel antiferromagnetic spintronic and valleytronic applications.
\end{abstract}
\maketitle
%
%
Antiferromagnets have many advantages for spintronic applications\cite{Baltz2018RMP, Jungwirth2016NatNano}, such as resistance to perturbing magnetic fields and the absence of stray fields.
For device operation, it is necessary that antiferromagnets exhibit clear magnetotransport responses like ferromagnets (e.g. magnetoresistance and Hall effects), although they do not in principle have spontaneous magnetisation\cite{Smejkal2022NatRevMater}.
Recently, large anomalous Hall and Nernst effects comparable to ferromagnets were discovered in several antiferromagnets in spite of the negligibly small spontaneous magnetisation\cite{Nakatsuji2015Nature, Nayak2016SciAdv, Ikhlas2018NatPhys, Ghimire2018NatCom, Park2022npgQM, Takagi2023NatPhys, Feng2022NatEle, Reichlova2024NatCom}.
Their electrically switchable properties have garnered significant attention, both for their potential applications and for their relevance to fundamental physics\cite{Higo2022Nature, Yoon2023NatMater, Han2024NatPhys}.
Notable examples include non-collinear magnets hosting cluster multipoles\cite{Nakatsuji2015Nature, Nayak2016SciAdv, Ikhlas2018NatPhys, Ghimire2018NatCom, Park2022npgQM, Takagi2023NatPhys}, as well as the so-called altermagnets\cite{Feng2022NatEle, Reichlova2024NatCom}.
A defining characteristic of these antiferromagnets is that their time-reversal ($\mathcal{T}$) symmetry is intrinsically broken due to their unique magnetic and crystal structures.
This causes spin splitting in the band structure, which is the source of their distinct magnetotransport phenomena.
\par
In those antiferromagnets, the parity ($\mathcal{P}$) symmetry associated with the spatial inversion is usually preserved.
However, when the $\mathcal{P}$-symmetry is also broken by the magnetic order, the band structure undergoes significant changes.
In such cases, both the $\mathcal{P}$- and $\mathcal{T}$-symmetries are broken, while the product $\mathcal{PT}$-symmetry is preserved.
The impact of $\mathcal{PT}$-symmetry on band structure was initially discussed for insulating magnetoelectric materials, such as Cr$_2$O$_3$, and has recently been extended to metallic magnets\cite{Wadley2016Science}.
Based on the electric/magnetic multipole theory, the $\mathcal{PT}$-symmetric magnetic order is characterised by the odd-parity magnetic multipole order\cite{Watanabe2024JPhysC, Hayami2024JPSJ}, resulting in a spin-degenerate band structure that is asymmetric in the momentum space\cite{Fedchenko2022JPhysC, Lytvynenko2023PRB}. 
This unique band structure was theoretically predicted to give rise to various intriguing transport and optical phenomena, including nonlinear electrical conductivity and photocurrent generation~\cite{Gao2014PRL, Zhang2019NatCom, Holder2020PRR, Fei2020PRB, Ahn2020PRX, Watanabe2018PRB, Watanabe2020PRR}.
While a few related experimental observations, such as nonlinear Hall effects~\cite{Godinho2018NatCom, Gao2023Science, Wang2023Nature} and bulk photogalvanic effects~\cite{Takagi2024APL}, have been reported, more direct manifestations of the asymmetric band structure have yet to be fully investigated.
One such phenomenon is the current-induced shift of asymmetric bands, which breaks the original symmetry and results in nematic distortion of the Fermi surface\cite{Watanabe2017PRB}.
This electronic deformation can generate the lattice deformation via the electron-phonon coupling, which is a magnetic and metallic counterpart of piezoelectricity (called magnetopiezoelectric effect) \cite{Watanabe2017PRB, Chazono2022PRB, Varjas2016PRL}.
In fact, it was experimentally observed that a very small change in sample length was caused by current in several $\mathcal{PT}$-symmetric antiferromagnets~\cite{Shiomi2018PRL, Shiomi2019PRB, Shiomi2020SciRep}.
However, intrinsic electronic responses (e.g., transport or optical changes) directly linked to the nematic deformation of the Fermi surface remain experimentally unexplored.
%
\begin{figure}
\begin{center}
\includegraphics[width=\linewidth]{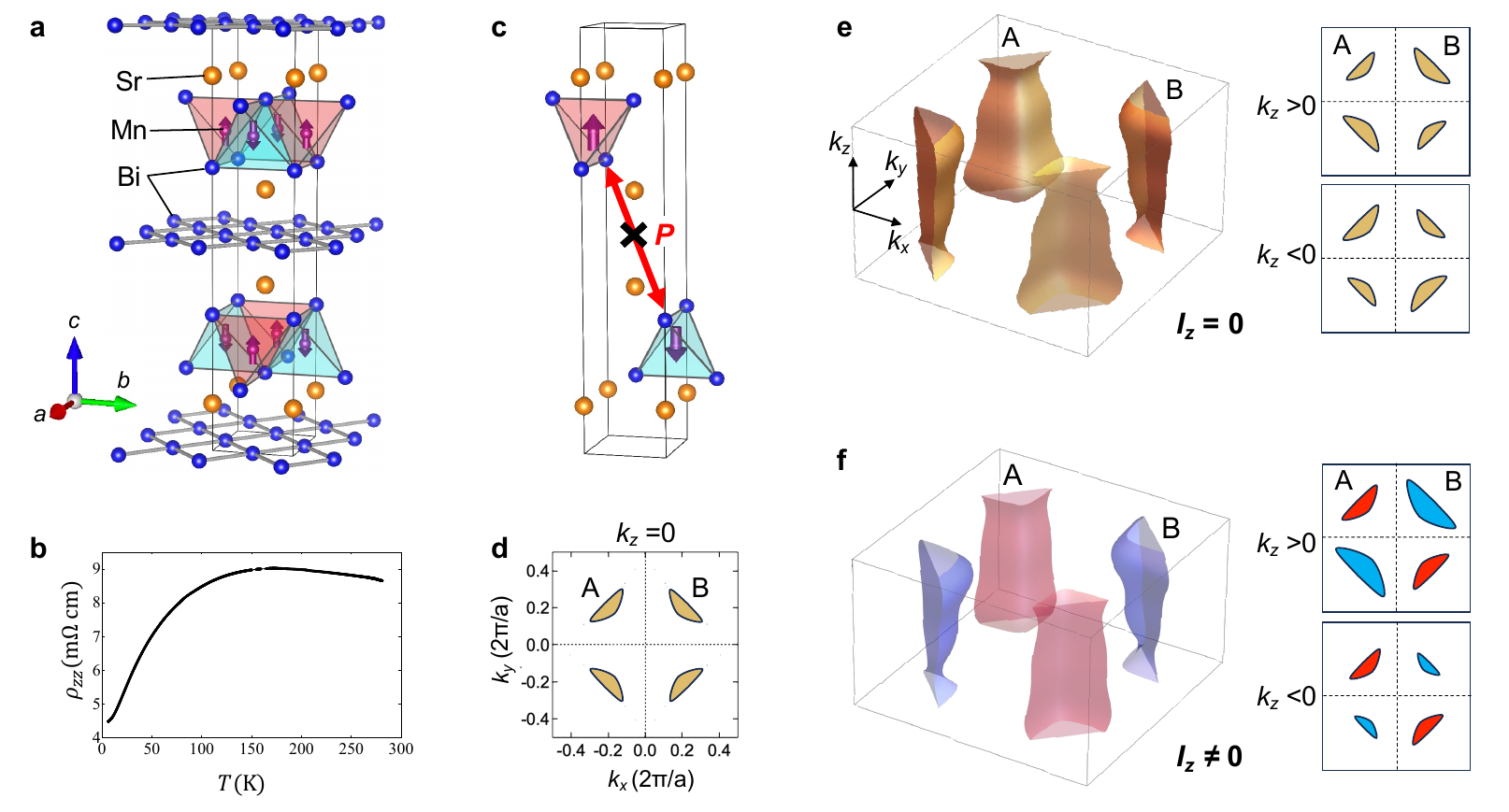}
\caption{%
\textbf{$\mathcal{PT}$-symmetric antiferromagnetic order and Fermi surfaces for SrMnBi$_2$.}
\textbf{a,} Lattice and magnetic structure of SrMnBi$_2$.
\textbf{b,} Temperature dependence of interlayer resistivity ($\rho_{zz}$) for a micro-fabricated device of SrMnBi$_2$.
\textbf{c,} Parity ($\mathcal{P}$) violation of antiferromagnetic order of the Mn sublattice.
\textbf{d,} Four-fold elliptic Dirac valleys at $k_z=0$. Valley A is elongated along $[110]$, while valley B along $[\overline{1}10]$.
\textbf{e,} Three-dimensional schematic Fermi surfaces (left) and corresponding cross-sectional cuts (right) for $k_z>0$ and $k_z<0$ of SrMnBi$_2$ with spin-orbit coupling for $I_z$=0.
($I_z$ is current along the $z$ direction.)
The asymmetry in $k_z$ dispersion is opposite for valleys A and B, preserving the overall four-fold symmetry.
\textbf{f.} Fermi surfaces for $I_z\ne0$.
 The current-induced modulation of asymmetric $k_z$ dispersion differs for each valley, breaking the original four-fold symmetry across the entire $k_z$-space.
}%
\end{center}
\end{figure}
\par
In this study, we have investigated transport phenomena associated with current-induced nematic deformation in the layered $\mathcal{PT}$-symmetric antiferromagnet SrMnBi$_2$~\cite{Park2011PRL, Wang2011PRB, WangPetrovic2011PRB}.
In this material, the local Mn spin surrounded by the Bi tetrahedron exhibits a staggered antiferromagnetic order\cite{Guo2014PRB}.
Therefore, the spin-up Mn is located at the centre of the red-coloured Bi tetrahedron, while the spin-down Mn is located at the centre of the blue-coloured Bi tetrahedron (Fig. 1a).
This leads to the simultaneous breaking of the $\mathcal{P}$- and $\mathcal{T}$-symmetries in a nonpolar manner, as the up-spin state in the red tetrahedron and the down-spin state in the blue tetrahedron are not related by either $\mathcal{P}$- or $\mathcal{T}$-operations (Figs. 1a and 1c).
Although this Mn-Bi layer is a Mott insulator, SrMnBi$_2$ also has a highly-conductive Bi square net layer sandwiched between the Mn-Bi layers, forming quasi-2D (gapped) Dirac bands\cite{Lee2013PRB, Feng2014SR, JiaPRB2014, Sakai2022JPSJreview}.
Since small amount of hole-type carriers are naturally doped, these bands allow the metallic conduction not only in the in-plane direction, but also in the out-of-plane direction by forming coherent cylindrical Fermi surfaces at low temperatures (see the metallic interlayer resistivity below $\sim$150 K in Fig. 1b)~\cite{Jo2014PRL, Kondo2020JPSJ}.
Therefore, SrMnBi$_2$ provides an ideal arena to study the current effects on a $\mathcal{PT}$-symmetric antiferromagnet\cite{Shiomi2018PRL, Shiomi2020SciRep}.
Furthermore, the Dirac bands in this material possess valley degrees of freedom ($vide$ $infra$), enabling unconventional valley symmetry breaking in conjunction with $\mathcal{PT}$-symmetric antiferromagnetic order.
\par
We here concretely explain the asymmetric nature of the quasi-2D Dirac bands in SrMnBi$_2$.
On the $k_z=0$ plane, these bands form exactly four-fold elliptic valleys (Fig. 1d)\cite{Lee2013PRB, Feng2014SR, JiaPRB2014}.
However, the $k_z$ dispersion, namely the warping of the cylindrical Fermi surface, is asymmetric with respect to $k_z=0$ due to the antisymmetric spin-orbit coupling arising from the odd-parity multipole order in the Mn-Bi layer.
First-principles calculations show that the cylindrical Fermi surface in the $k_z>0$ ($k_z<0$) regime contracts (expands) for the valley A elongated along $[110]$ direction, while the opposite $k_z$ dispersion occurs for the valley B elongated along the $[1\overline{1}0]$ direction, as schematically shown in Fig. 1e (For the quantitative $k_z$ dispersion, see Fig. S1).
When an electric current $I_z$ is applied along the $z$ direction, such asymmetrically warped cylindrical Fermi surfaces shift along the $z$ direction, leading to the four-fold symmetry breaking.
More specifically, the valleys A and B are non-equivalent in terms of size and anisotropy across the entire $k_z$ for nonzero $I_z$ (Fig. 1f).
%
\begin{figure}
\begin{center}
\includegraphics[width=\linewidth]{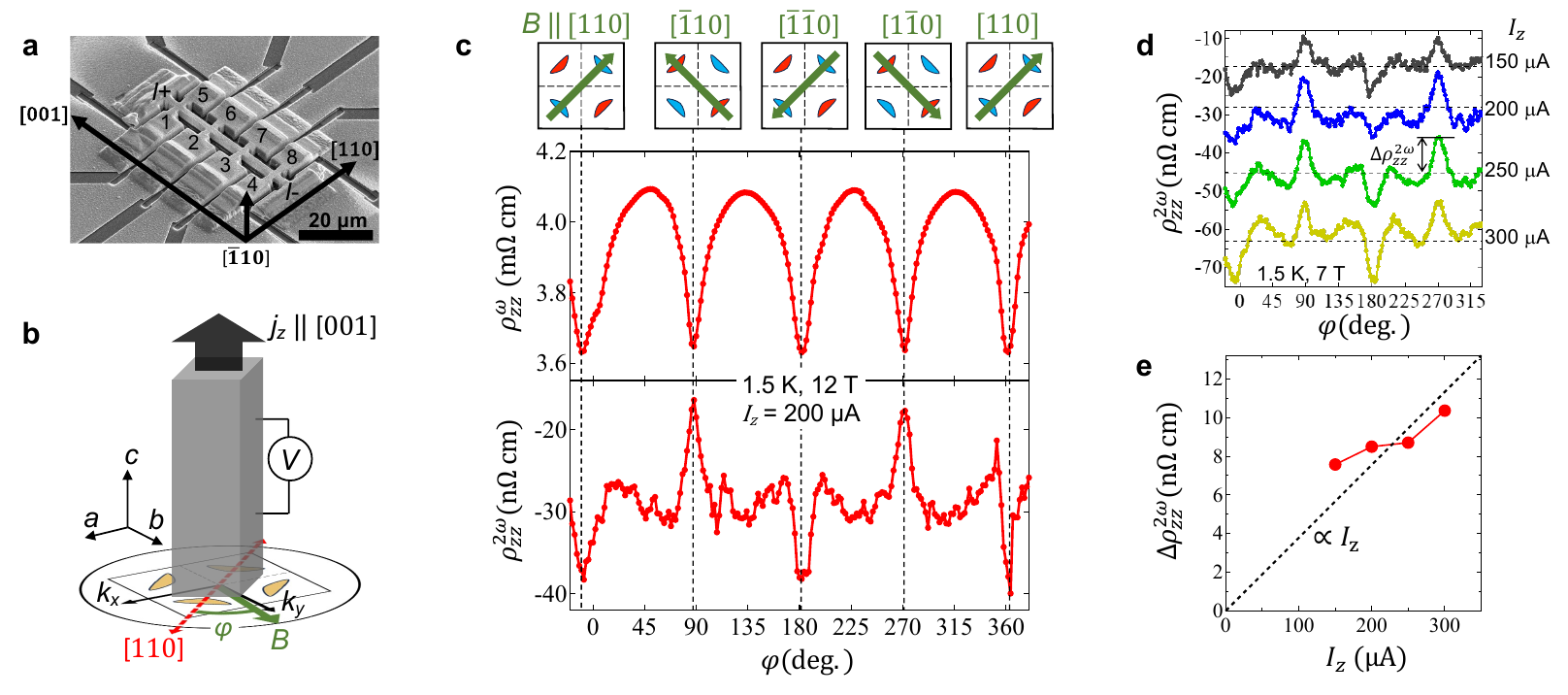}
\caption{%
\textbf{Nonreciprocal angular magnetoresistance effects in SrMnBi$_2$}
\textbf{a,} SEM image of the micro-fabricated device of SrMnBi$_2$.
\textbf{b,} Schematic illustration of the measurement of angular magnetoresistance effects of interlayer resistivity for the field rotating within the $ab$ plane.
\textbf{c,} Azimuth-angle $\varphi$ dependence of the first-harmonic interlayer resistivity $\rho^\omega_{zz}$ (upper) and the second-harmonic one $\rho^{2\omega}_{zz}$ (lower) at 1.5 K at 12 T, measured between the terminals 2-3. The magnitude and frequency of the applied current $I_z$ were 200 $\mu$A and 23 Hz, respectively. The inset shows the relation between the field direction and Dirac valleys.
\textbf{d,} $\rho^{2\omega}_{zz}$ versus $\varphi$ at 1.4 K at 12 T for various $I_z$. The dotted horizontal line is the base line for each curve, which is determined as the average of the peak and dip values. Each curve is shifted for clarity.
\textbf{e,} Peak height of $\rho^{2\omega}_{zz}$ ($\Delta \rho^{2\omega}_{zz}$ defined in \textbf{d}) versus $I_z$. 
}
\end{center}
\end{figure}
\par
For revealing an impact of such current-induced nematic deformation of Dirac valleys on transport phenomenon, angular-dependent magnetoresistance (AMR) measured by rotating the field within the $ab$ plane is promising.
Owing to the strong anisotropy of each Dirac valley in this material, the interlayer resistivity exhibits large four-fold AMR with sharp angle dependence (Fig. 2b)\cite{Jo2014PRL, Kondo2020JPSJ}, which is of great advantage to detect broken four-fold symmetry.
Considering that the deformation of valleys is induced by current, the change in AMR should be proportional to current density, which results in a nonreciprocal (i.e., diode-like) effect on the interlayer resistivity.
To detect this, we measured the second-harmonic component of voltage drop between the terminals as well as the conventional first-harmonic component.
\par
To achieve high current density without significant current heating effects, we fabricated micro-devices (Fig. 2a) capable of reaching a current density of $j_z\sim5\times10^7$ A/m$^2$ with an applied current of $I_z=200$ $\mu$A.
Figure 2c shows the AMR profile (i.e., azimuth angle $\varphi$ dependence) of the first-harmonic component of interlayer resistivity $\rho^{\omega}_{zz}$ at 1.5 K at 12 T.
As a result of precise alignment of the field perpendicular to the $c$ axis using a two-axis rotation probe, we observed the four-fold $\varphi$ dependence of $\rho^\omega_{zz}$.
There, $\rho_{zz}^{\omega}(\varphi)$ shows narrow dips for $\bm{B}\parallel[110]$ and $[\overline{1}10]$ ($\varphi=0^\circ, 90^\circ, \cdots$) while it shows broad peaks for $\bm{B}\parallel[100]$ and $[010]$ ($\varphi=45^\circ, 135^\circ, \cdots$), as is consistent with the literature\cite{Jo2014PRL, Kondo2020JPSJ}.
Since the direction of the nematic deformation changes by $90^\circ$ depending on the sign of $j_z$, its impact on $\rho^{\omega}_{zz}$ is averaged out and the four-fold symmetry is preserved.
On the other hand, the second-harmonic component of interlayer resistivity $\rho^{2\omega}_{zz}(\varphi)$ exhibits distinct two-fold symmetry: sharp dips are located at $\varphi=0^\circ, 180^\circ$ ($\bm{B}\parallel[110]$, $[\overline{1}\overline{1}0]$), while sharp peaks are located at $\varphi=90^\circ, 270^\circ$ ($\bm{B}\parallel[\overline{1}10]$, $[1\overline{1}0]$).
This clearly indicates that the valleys A and B are non-equivalent when $j_z$ is applied.
Reflecting that the deformation of each valley is driven by current, the peak height of $\rho^{2\omega}_{zz}$ (denoted by $\Delta \rho^{2\omega}_{zz}$ in Fig. 2d) is almost proportional to $j_z$ (Fig. 2e).
%
\begin{figure}
\begin{center}
\includegraphics[width=\linewidth]{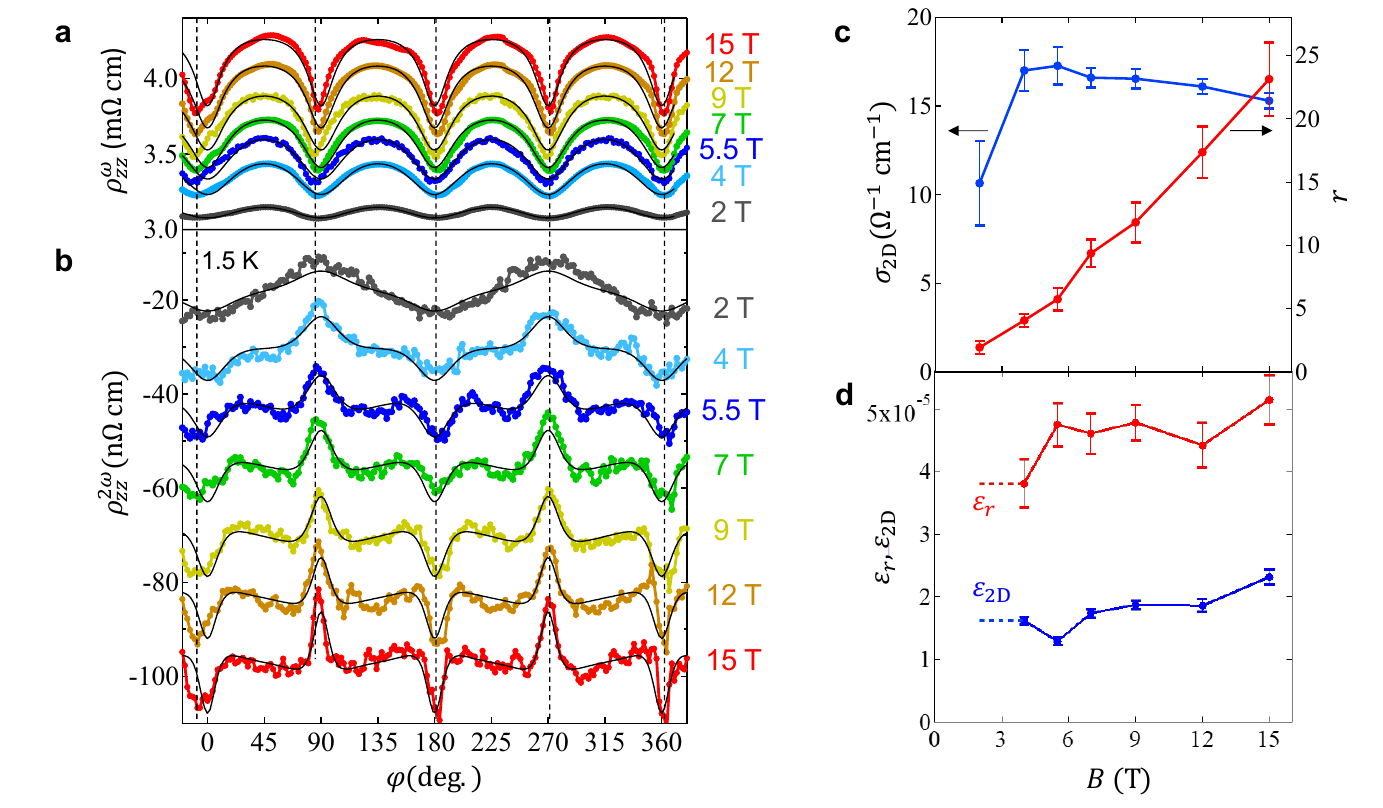}
\caption{
\textbf{Magnetic-field dependence and fitted results.}
\textbf{a, b} $\varphi$ dependences of $\rho^\omega_{zz}$ (\textbf{a}) and $\rho^{2\omega}_{zz}$ (\textbf{b}) for SrMnBi$_2$ at 1.5 K for $I_z$=200 $\mu$A for various magnetic fields. The solid curves are the fitted results of the experimental data on the basis of the empirical equation (see the main text).
\textbf{c, d} Magnetic-field $B$ dependence of the fitted parameters. $\sigma_{\rm 2D}$ (\textbf{c}) and $r$ (\textbf{c}) are parameters for $\rho^\omega_{zz}$. $\varepsilon_{\rm 2D}$ (\textbf{d}) and $\varepsilon_r$  (\textbf{d}) are the current-induced variations of $\sigma_{\rm 2D}$ and $r$, respectively, determined by fitting $\rho^{2\omega}_{zz}$ (see also the main text).
}%
\end{center}
\end{figure}
\par
The profile of $\rho^{2\omega}_{zz}(\varphi)$ strongly depends on magnetic field (Fig. 3b).
At 2 T, it exhibits a sine-like curve with two-fold symmetry.
However, with increasing field, the dip (peak) structures located at $\varphi=0^\circ, 180^\circ$ ($\varphi=90^\circ, 270^\circ$) progressively evolve, accompanied by the gradual change in the weak background component.
As discussed below, the observed variation in $\rho^{2\omega}_{zz}(\varphi)$ is explained by the field dependence of $\rho^\omega_{zz}(\varphi)$; the dips of $\rho^\omega_{zz}(\varphi)$ (at $\varphi$=0$^\circ$, 90$^\circ$, 180$^\circ$, $\cdots$) become deeper and narrower with increasing field (Fig. 3a).
\par
To formulate the relation between $\rho^{2\omega}_{zz}(\varphi)$ and $\rho^{\omega}_{zz}(\varphi)$, we here employ a phenomenological model of interlayer magnetoconductivity $\sigma_{zz}\,(=1/\rho^\omega_{zz})$ taking account of the in-plane anisotropy of quasi-2D Dirac valley (Fig. 1d)~\cite{Jo2014PRL, Kondo2020JPSJ, Zhu2012NatPhys}:
\begin{equation}\label{eq:AMR}
	\sigma_{zz}(\varphi)=\frac{2\sigma_{\rm 2D}}{1+r{\cos}^{2}\varphi}+\frac{2\sigma_{\rm 2D}}{1+r\cos^{2}(\varphi+\pi/2)}+\sigma_{3D},
\end{equation}
where $\sigma_{\rm 2D}$ ($\sigma_{\rm 3D}$) is the relative contribution of each quasi-2D Dirac valley (all 3D Fermi surfaces from the parabolic bands)\cite{Jo2014PRL, Kondo2020JPSJ}.  
$r$ is the anisotropic factor of magnetoconductivity, resulting in the maximum (minimum) conductivity for the field along the shorter (longer) axis of the elliptic Dirac valley.
Note here that the first (second) term corresponds to the contribution from the valley A (B), which gives $\sigma_{zz}$ peaks, i.e., $\rho^\omega_{zz}$ dips at $\varphi = 90^{\circ}, 270^{\circ}$ ($\varphi = 0^{\circ}, 180^{\circ}$).
The experimental profiles of $\rho^\omega_{zz}(\varphi)$ at various magnetic fields are nicely fitted by Eq. \ref{eq:AMR}, as denoted by solid curves in Fig. 3a.
The fitted values of $r$ and $\sigma_{\rm 2D}$ are summarised in Fig. 3c.
Reflecting the deeper and narrower dips in $\rho^\omega_{zz}(\varphi)$ at higher fields, the $r$ value monotonically increases with increasing field, whereas the $\sigma_{\rm 2D}$ is almost independent of field (the same is true for $\sigma_{\rm 3D}$, see Fig. S2).
These fitted results are consistent with those reported in the literatures~\cite{Jo2014PRL, Kondo2020JPSJ}.
\par
We now take account of the impact of the current-induced nematicity on Eq. \ref{eq:AMR}.
Considering that the valleys A and B are non-equivalent in the presence of $j_z$, the nematicity can be incorporated as current-induced changes in $r$ and $\sigma_{\rm 2D}$ as follows
\begin{eqnarray}\label{eq:epsilon}
		r&\rightarrow& r(1\pm\epsilon_r),\\
		\sigma_{\rm 2D}&\rightarrow& \sigma_{\rm 2D}(1\pm\epsilon_{\rm 2D})
\end{eqnarray}
where  $\epsilon_r\ (\propto j_z)$ and $\epsilon_{\rm 2D}\ (\propto j_z)$ are dimensionless variations in $r$ and $\sigma_{\rm 2D}$, respectively.
Note here that the $+$ sign corresponds to the first term (valley A) in Eq. \ref{eq:AMR}, while the $-$ sign corresponds to the second term (valley B) in Eq. \ref{eq:AMR}.
The resultant variation in $\sigma_{zz}$ is given to the first order of $\epsilon_r$ and $\epsilon_{\rm 2D}$ by (see Methods)
\begin{equation}\label{eq:sigma_1w_2w}
			\sigma_{zz}(\varphi)\rightarrow\sigma_{zz}(\varphi)+\delta\sigma_{zz}(\varphi),
\end{equation}
where 
\begin{equation}\label{eq:sigma_2w}
	\begin{split}
		\delta\sigma_{zz}(\varphi)
		&=-2\sigma_{\rm 2D}r\left\lbrace \left( \frac{\rm{cos}\varphi}{1+r\rm{cos}^{2}\varphi}\right) ^2-\left( \frac{\rm{cos}(\varphi+\pi/2)}{1+r\rm{cos}^{2}(\varphi+\pi/2)}\right) ^2\right\rbrace\epsilon_r\\
		&\qquad+2\sigma_{\rm 2D}\left\lbrace\frac{1}{1+r\rm{cos}^{2}\varphi}-\frac{1}{1+r\rm{cos}^{2}(\varphi+\pi/2)} \right\rbrace\epsilon_{\rm 2D} \\
	\end{split}
\end{equation}\\
Since $\epsilon_r$ and $\epsilon_{\rm 2D}$ are proportional to $j_z$, $\delta\sigma_{zz}$ is also proportional to $j_z$, corresponding to the nonreciprocal component of the interlayer conductivity.
By converting the nonreciprocal conductivity to resistivity using the relation $\rho_{zz}^{2\omega} = -\frac{\delta\sigma_{zz}}{(\sigma_{zz})^2}$ (see Methods), we fit the experimental $\rho^{2\omega}_{zz}(\varphi)$ profiles at various fields, as shown in Fig.~3b. 
For this fitting, $\epsilon_{\rm 2D}$ and $\epsilon_r$ were treated as adjustable parameters, while $\sigma_{\rm 2D}$ and $r$ were fixed to the values obtained from fitting $\rho_{zz}^\omega(\varphi)$ at each field (Fig.~3c). 
The experimental $\rho^{2\omega}_{zz}(\varphi)$ profiles are well reproduced by Eq.~\ref{eq:sigma_2w} regardless of field, as indicated by the solid curves in Fig.~3b.
This demonstrates that the field dependence of $\rho^{2\omega}_{zz}(\varphi)$ arises solely from the field dependence of $\rho^{\omega}_{zz}(\varphi)$.
The extracted values of $\epsilon_{\rm 2D}$ and $\epsilon_r$ are nearly field-independent and are of the order of $10^{-5}$ (Fig.~3d).
Consequently, the magnitude of current-induced nematicity is estimated to be of the order of $10^{-5}$ for $j_z\sim5\times10^7$ A/m$^2$ ($I_z$=200 $\mu$A).
In the previous study, current-induced lattice displacement was measured in EuMnBi$_2$, where the largest strain was $\sim1.4\times 10^{-8}$ at $j_z\sim4\times10^4$ A/m$^2$ at the lowest temperature~\cite{Shiomi2020SciRep}.
This corresponds to the strain of $\sim2\times10^{-5}$ at $j_z\sim5\times10^7$ A/m$^2$ (the same $j_z$ used in this study).
Thus, the nematicity estimated from the AMR profile is of the same order of magnitude as that revealed by the lattice displacement measurements.
%
\begin{figure}
\begin{center}
\includegraphics[width=\linewidth]{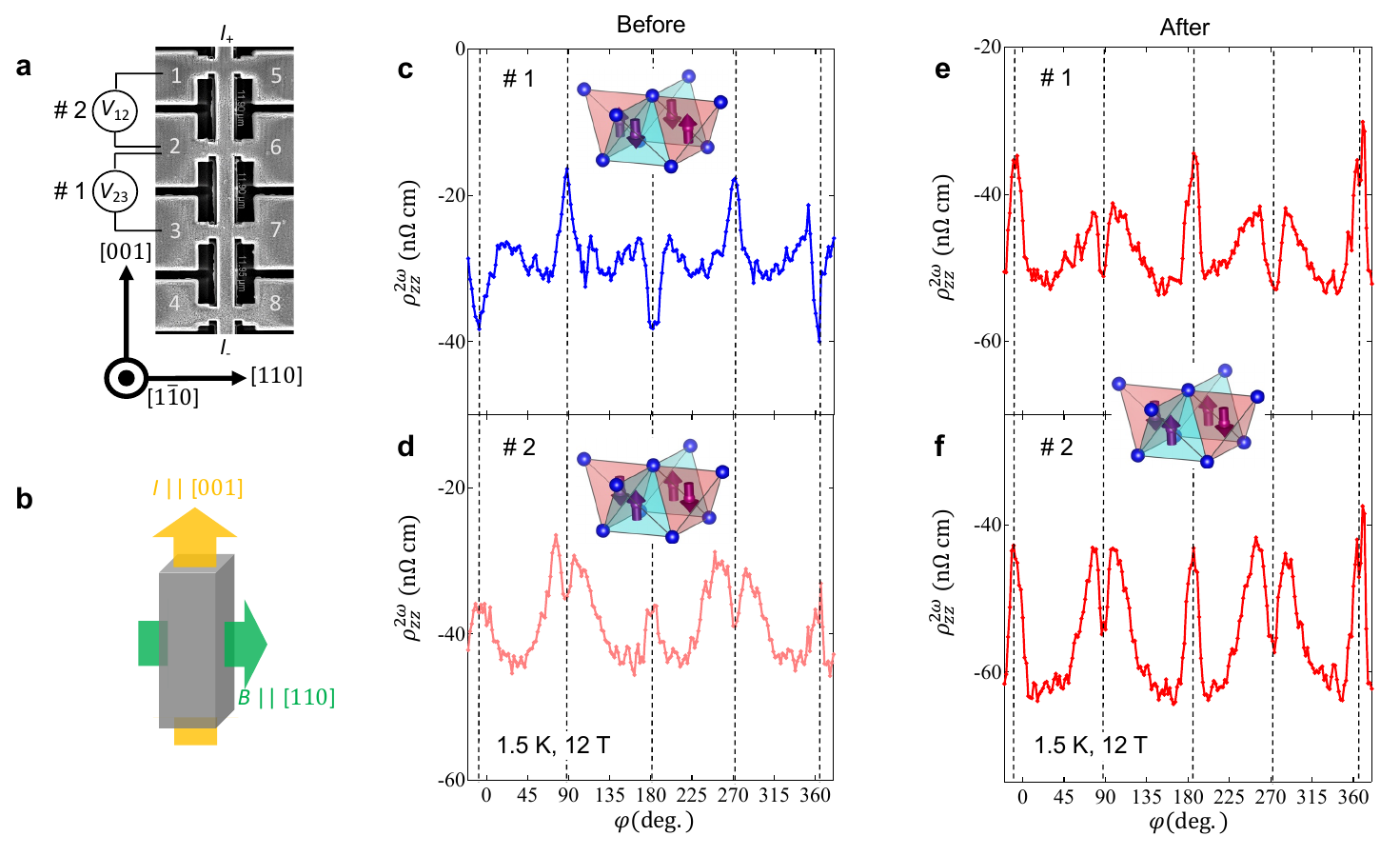}
\caption{
\textbf{Electric-magnetic control of the nonpolar $\mathcal{PT}$-symmetric antiferromagnetic order.}
\textbf{a} SEM image of the neighbouring devices. The interlayer resistivity in device \#1 was measured between terminals 2-3, while in device \#2, it was measured between terminals 1-2.
\textbf{b} Schematic showing the direction of the electric current ($I$) along [001] and the magnetic field ($B$) along [110] for domain poling.
\textbf{c, d} $\varphi$ dependence of $\rho_{zz}^{2\omega}(\phi)$ in devices \#1 (\textbf{c}) and \#2 (\textbf{d}) at 1.5 K and 12 T for $I_z = 200$ $\mu$A before domain poling.
\textbf{e, f}  Corresponding $\rho_{zz}^{2\omega}(\phi)$ profiles for devices \#1 (\textbf{e}) and \#2 (\textbf{f}) after domain poling by current-field cooling, showing an inversion of the peak and dip positions in device \#1.
The insets illustrate the domain of antiferromagnetic order in the Mn-Bi layers of each device. 
}%
\end{center}
\end{figure}
\par
Finally, we demonstrate the detection and control of magnetic domains in the present $\mathcal{PT}$-symmetric antiferromagnet.
Figures 4c and 4d show the profiles of $\rho^{2\omega}_{zz}(\varphi)$ for the two adjacent devices \#1 and \#2, respectively, after cooling from room temperature without applying any field or current.
Notably, the positions of the sharp peaks and dips of $\rho^{2\omega}_{zz}(\varphi)$ are reversed between \#1 and \#2, apart from the relatively large background component in \#2~\cite{BG_2nd_device}.
This clearly indicates that the sign of $\epsilon_r$ and $\epsilon_{\rm 2D}$, i.e., the domain of antiferromagnetic order, differs between the two devices, even within the same micro-fabricated crystal (inset to Figs. 4c and 4d).
\par
To align the domains, we employed a poling procedure conducted above the N\'{e}el temperature $T_{\rm N}$=295 K~\cite{Guo2014PRB}.
Symmetry analysis predicts that the domains of $\mathcal{PT}$-symmetric antiferromagnetic order in this nonpolar system can be aligned by applying an electric current and a magnetic field orthogonal to each other, as illustrated in Fig.~4b (see discussions below)\cite{Watanabe2018PRBR}.
To achieve this experimentally, we applied a current of 1~mA ($j_z \sim 3 \times 10^8~\mathrm{A/m^2}$) along [001] and a field of 18~T along [110] at 305~K ($>T_{\rm N}$), and then cooled the sample to 4.2~K.
After poling, we repeated the AMR measurements at 1.5~K and 12~T.
In device \#1, the peak and dip of $\rho^{2\omega}_{zz}(\varphi)$ are inverted compared to the previous measurement without poling (Fig. 4e).
On the other hand, in device \#2,  the magnitude of the peak and dip of $\rho^{2\omega}_{zz}(\varphi)$ is enhanced while keeping their positions (Fig. 4f).
These results indicate that the antiferromagnetic domains in both devices were aligned to the original domain in device \#2 through the poling procedure, thereby demonstrating electric-magnetic manipulation of the nonpolar $\mathcal{PT}$-symmetric antiferromagnetic order.
\par
We here discuss the mechanisms underlying this control of antiferromagnetic domains in the present material.
Recently, several current-driven domain control methods have been developed for metallic antiferromagnets, although their variety remains limited~\cite{Jiang2020NatCom,Masuda2024NatCom}.
For $\mathcal{T}$-odd antiferromagnets with spin-split bands~\cite{Smejkal2022NatRevMater}, conventional control mechanisms, similar to those in ferromagnets, can be employed. 
Examples include spin-orbit torque switching at the interfaces of artificial heterostructures~\cite{Higo2022Nature, Yoon2023NatMater, Han2024NatPhys}.
In contrast, the control mechanisms for $\mathcal{PT}$-symmetric antiferromagnets with spin-degenerate bands are significantly more intricate.
In the polar systems, such as CuMnAs~\cite{Wadley2016Science, Godinho2018NatCom} and Mn$_2$Au~\cite{Bodnar2018NatCom}, current-induced switching was achieved via the sublattice-dependent spin-momentum locking~\cite{Zelezny2014PRL, Yanase2014JPSJ}.
However, in nonpolar systems like SrMnBi$_2$, the absence of polarity ($p$-wave) in momentum space prevents the domain switching using current alone.
Instead, the higher-order $f$-wave polarity in momentum space can be utilised.
To this end, we have applied an in-plane field, effectively rendering the system polar and enabling current-induced switching.
This combined electric and magnetic manipulation of $\mathcal{PT}$-symmetric antiferromagnetic domains provides a novel protocol distinct from those employed for $\mathcal{T}$-odd or polar antiferromagnets currently under active investigation.
\par
To conclude, we report the observation of current-induced electronic nematicity in the nonpolar $\mathcal{PT}$-symmetric antiferromagnet SrMnBi$_2$ by measuring the nonreciprocal angular magnetoresistance (AMR) effect.
The breaking of the original four-fold symmetry is detectable via the valley degrees of freedom of Dirac fermions in the Bi square net, which is adjacent to the Mn-Bi tetrahedral layer exhibiting $\mathcal{PT}$-symmetric antiferromagnetic order.
By employing a phenomenological model of the AMR effect attributed to elliptic Dirac valleys, we quantitatively reproduce the two-fold nonreciprocal AMR signal, demonstrating the current-induced lifting of the valley degeneracy.
The layered structure incorporating a Dirac fermion layer in this material offers a novel platform for the electrical control of valleys by the $\mathcal{PT}$-symmetric antiferromagnetic order.
Our findings may pave the way for developing unconventional spintronic and valleytronic devices, thereby broadening the scope of antiferromagnetic spintronics.
\begin{acknowledgments}
The authors thank Y. Onose and A. P. Mackenzie for helpful discussions.
This work was partly supported by the JSPS KAKENHI (Grants No. 22H00109, 22K18689, 23H00268, 23H04862, 23H04014, 23H04868, 23KK0052, 22H04933, 23K22447, 21H05470, and 24H01622), the Asahi Glass Foundation and the Spintronics Research Network of Japan (Spin-RNJ).
A part of this work was carried out under the Visiting Researcher's Program and the GIMRT Program of the Institute for Materials Research, Tohoku University.
\end{acknowledgments}
%
\section{Methods}
\subsection{Sample preparation and device fabrication}
Single crystals of SrMnBi$_2$ were grown by a Bi self-flux method described elsewhere\cite{Park2011PRL, Kondo2020JPSJ, Masuda2016SA}.
To measure the interlayer resistivity with a high current density, we fabricated microstructured devices using a focused ion beam (FIB) with 30 kV of the acceleration voltage for the Ga ion beam at VERSA-3D (FEI Company).
The typical dimension of the fabricated device is 2 $\mu$m $\times$ 2 $\mu$m $\times$ 12 $\mu$m with 8 voltage terminals and 2 current terminals.
\subsection{Transport measurement}
$\rho^{\omega}_{zz}$ and $\rho^{2\omega}_{zz}$ were measured by the conventional AC-lockin measurement with a typical frequency of 23 Hz.
$\rho^{2\omega}_{zz}$ was obtained as $\rho^{2\omega}_{zz}=(V^{2\omega}_{zz}/I_z)/(l/S)$, where $V^{2\omega}_{zz}$ is the imaginary part of the second-harmonic voltage drop, $I_z$ is the RMS value of AC current, $l$ is the length of voltage terminals, and $S$ is the cross section of current path.
To obtain the $B$ dependence, we symmetrize $V^{2\omega}_{zz}$ with respect to $B$.
The measurements were performed using commercial superconducting magnets, where the direction of magnetic field was controlled using a probe equipped with a two-axis rotating stage in Institute for Materials Research, Tohoku University.
\subsection{Formulation of the fitting function of $\rho^{2\omega}_{zz}$}
Considering the current-induced nematicity described by Eq.~\ref{eq:epsilon}, the variation of the first term (associated with valley A) in Eq.~\ref{eq:AMR} can be expressed, to the first order of $\epsilon_r$ and $\epsilon_{\rm 2D}$, as:
\begin{equation}\label{eq:rA_j}
	\begin{split}
		\frac{2\sigma_{\rm 2D}}{1+r\rm{cos}^{2}\varphi}&\rightarrow\frac{2\sigma_{\rm 2D}(1+\epsilon_{\rm 2D})}{1+r(1+\epsilon_r)\rm{cos}^{2}\varphi}\\
		&\sim\frac{2\sigma_{\rm 2D}}{1+r\rm{cos}^{2}\varphi}-2\sigma_{\rm 2D}r\left(\frac{\rm{cos}\varphi}{1+r\rm{cos}^{2}\varphi}\right)^2\epsilon_r+2\sigma_{\rm 2D}\frac{1}{1+r\rm{cos}^{2}\varphi}\epsilon_{\rm 2D}.\\
	\end{split}
\end{equation}\\
The second and third terms in the above equation originate from the current-proportional variations in $r$ and $\sigma_{\rm 2D}$, respectively. Similarly, the variation of the second term in Eq.~\ref{eq:AMR} can be derived to the first order of $\epsilon_r$ and $\epsilon_{\rm 2D}$, resulting in the total variation given by $\delta\sigma_{zz}(\varphi)$ in Eq.~\ref{eq:sigma_2w}.
\par
The relation between the current density $j_z$ and the applied electric field $E_z$ is expressed as:
\begin{equation}
j_z=(\sigma_{zz}+\delta\sigma_{zz})E_z.
\end{equation}
From this, we obtain
\begin{align}
E_z &= \frac{j_z}{\sigma_{zz} + \delta\sigma_{zz}} \notag \\
&\sim \left(\frac{1}{\sigma_{zz}} - \frac{\delta\sigma_{zz}}{\sigma_{zz}^2}\right)j_z,
\end{align}
where $\frac{1}{\sigma_{zz}}$ and $-\frac{\delta\sigma_{zz}}{\sigma_{zz}^2}$ in the final equation correspond to $\rho^{\omega}_{zz}$ and $\rho^{2\omega}_{zz}$, respectively.
\subsection{First-principles calculation}
We performed first-principles band structure calculations based on the density functional theory with the generalised gradient approximation with the Perdew-Burke-Ernzerhof parametrization~\cite{PBE} and the projector augmented wave (PAW) method as implemented in the Vienna {\it ab initio} simulation package~\cite{paw, vasp1,vasp2,vasp3,vasp4}.
We used the experimental crystal structure\cite{Cordier1977ZNB} with the antiferromagnetic ordering shown in Fig. 1a.
Core-electron states in PAW potentials were [Ar]$3d^{10}$, [Ar], and [Xe]$5d^{10}$ for Sr, Mn, and Bi, respectively.
We used a plane-wave cutoff energy of 400 eV for Kohn-Sham orbitals with including the spin-orbit coupling.
A $16\times 16\times 16$ ${\bm k}$-mesh was used.
After first-principles band calculation, we extracted Mn-$d$ + Bi-$p$ Wannier orbitals using Wannier90 software~\cite{wan1, wan2, wan3}.
Fermi surfaces were depicted using the tight-binding model consisting of these Wannier orbitals.
%


\begin{thebibliography}{99}
\bibitem{BG_2nd_device} In the $\rho^{2\omega}_{zz}(\varphi)$ profile of device \#2, the gradual background (except for the sharp peak/dip structures) is more pronounced than in device \#1. As shown later, this background component does not change before and after the poling. Thus, it might arise from some pinned domains and/or other extrinsic effects.
\end{thebibliography}
\end{document}